\documentstyle[twocolumn,epsf,aps]{revtex}
\draft
\begin{document}
\title{Branching annihilating random walks with parity conservation\\
on a square lattice}
\author{Gy\"orgy Szab\'o$^{1,2}$ and Maria Augusta Santos$^2$}

\address
{$^1$ Research Institute for Technical Physics and Materials Science \\
POB 49, H-1525 Budapest, Hungary}

\address
{$^2$ Departamento de F\'{\i}sica and Centro de F\'\i sica do Porto \\
Faculdade de Ci\^encias, Universidade do Porto \\
Rua do Campo Alegre 687, 4150 Porto, Portugal}

\address{
\centering{
\medskip \em
\begin{minipage}{15.4cm}
{}\qquad Using Monte Carlo simulations we have studied the transition from an
``active'' steady state to an absorbing ``inactive'' state for two versions
of the branching annihilating random walks with parity conservation
on a square lattice. In the first model the randomly walking particles
annihilate when they meet and the branching process creates two 
additional particles; in the second case we distinguish
particles and antiparticles created and annihilated in pairs.
Quite distinct critical behavior is found in the two cases, raising the
question of what determines universality in this kind of systems.
\end{minipage}
}}
\maketitle

\narrowtext

Branching annihilating random walks (BARWs) have been extensively studied 
in recent years because they are a prototype for a variety of
reaction-diffusion-like systems (for recent reviews see the work by
Cardy and T\"auber \cite{CT98} and 
by Marro and Dickman \cite{MD} for more general aspects).
For different models, the random walkers can represent either domain walls
(kinks) or active sites on a lattice. In general, the corresponding
critical behavior belongs to the directed percolation (DP) universality
class. According to the "DP conjecture"
\cite{dpconj}, most of the one-component models with a single absorbing
state belong to the DP universality class. Exceptions can appear when
additional symmetries or conservation laws are introduced. The best
known one-dimensional (1D) exceptions are models in which, either the
parity of the number of particles is conserved during the elementary
processes \cite{grass} - \cite{marques}, or
there is an equivalence between two absorbing states \cite{KP,haye}.
Henceforth, we will concentrate on this type of 2D BARWs whose behavior
differs from the 2D DP universality class \cite{2dDP}.

A field theoretical analysis of such systems was recently reported
\cite{CT96,CT98}. In 1D, the numerical and theoretical approaches are
in satisfactory agreement: a new universality class appears when parity
is preserved. Agreement is also found  for $D > 2$ in which case the
mean-field results are valid. The nonexistence of the absorbing
state (extinction) for finite values of the branching rate predicted by
mean-field approximation for $D \ge 2$ was also confirmed by early Monte
Carlo (MC) simulations \cite{TT} in 2D and 3D. However, according to
the field theoretical analysis, logarithmic corrections are expected for
the marginal dimensionality $D=2$ for models with parity conservation
and this was not born out by previous work. In this communication we
report results of extensive MC simulations in 2D lattices showing evidence
of such corrections. Inspired by the kink/antikink interpretation of creation
and annihilation, we have also considered a modification of the model
where two types of particles are present in the creation and annihilation
processes and indeed found that a distinct behavior appears in this case.

In the first model there is only one species of particles
walking randomly on a square lattice. In addition to the single particle
diffusion, the time evolution is governed by creation and annihilation
of particle pairs as follows. A randomly chosen particle creates two
additional particles with a probability $p$ which are located on
two randomly chosen neighboring empty sites \cite{notation}; otherwise the
chosen particle jumps to one of its nearest neighbor positions. For both 
elementary processes, if the destination site is occupied then the two
particles annihilate. In order to study extinction, we consider only
even initial numbers of particles.

In the second model we distinguish particles and antiparticles, the
creation and annihilation processes involving a particle/antiparticle
pair with an evolution rule similar
to the above one. During the sequential updating we neglect all the
elementary processes which would result in two particles
(or antiparticles) on the same site. The numbers of particles and
antiparticles are chosen to be equal. As a result we have a
unique absorbing state (no particles) independent of time. This model
can be considered as a 2D generalization of the parity conserving model
introduced by Menyh\'ard \cite{MO} where the 1D
ferromagnetic domain walls are represented by particles and
antiparticles. It is worth mentioning that in the former model the
particles and antiparticles are alternately positioned along the chain and
this feature is maintained by the elementary processes, therefore
their distinction is not relevant. In the present 2D model, however,
two particles (antiparticles) can occupy neighboring positions and
they can avoid each other.

The MC simulations are performed on an $L \times L$ square lattice with
periodic boundary conditions for different values of the branching rate
$p$. In order to have sufficiently accurate results, the system size
($L$) is increased up to $L=2000$ for small concentration of particles.
The simulations are started from a randomly half-filled lattice and
during the evolution we record the concentration of particles (other 
initial conditions were also tested).
Time is measured in Monte Carlo steps (MCS) within which each
particle has an opportunity to jump (probability $1-p$) or
branch (probability $p$).

For both models, in the absence of branching ($p=0$), the number of
particles decreases monotonously and eventually vanishes. For $p > 0$,
however, the system remains active with
a fluctuating number of particles $n$ if the lattice size is sufficiently
large. In the stationary state the average concentration
($c=\langle n \rangle / L^2$) of particles vanishes continuously when $p$ 
tends to $0$. In other words, decreasing the branching rate,
both systems undergo a transition with a critical point $p_c=0$.

First we have investigated the decrease of concentration at the critical
point ($p_c=0$). For this purpose the time-dependent average concentration
and its fluctuation \mbox {$\chi = L^2 \langle (c - n/L^2)^2 \rangle $}
are determined at discrete time steps (equidistant in the logarithmic
scale) by averaging over 500 runs.

\begin{figure}
\centerline{\epsfxsize=7.5cm
            \epsfbox{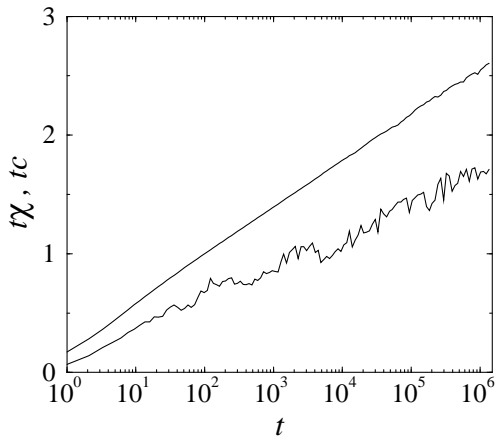}
            \vspace*{2mm}   }
\caption{Concentration (upper curve) and its fluctuation (lower curve)
multiplied by the time {\it vs.} the logarithm of time at zero branching
rate for the first model.}
\label{fig:marw}
\end{figure}

For the first model the field theoretical investigation suggests
$c \propto \ln t /t $ \cite{CT98}. To check this prediction we
have plotted $t c$ {\it vs.} $t$ in a log-lin plot. Our MC results
(see Fig.~\ref{fig:marw}) obtained for $L=2000$ indicate clearly that
the time-dependent concentration can be well described as
\begin{equation}
c(t) = {A + B \ln t \over t }
\end{equation}
for sufficiently long times ($t > 100$ MCS). This function fits the
MC data if $A=0.2238$ and $B=0.8979$. The leading term of this
asymptotic behavior agrees with the above prediction given by Lee \cite{Lee}
and Cardy and T\"auber \cite{CT98}.

The fluctuation of concentration decreases proportionally with the
average value of concentration as demonstrated in Fig.~\ref{fig:marw}.
Neglecting the ``noisy decoration'' due to the statistical error,
the ratio of the time-dependent fluctuation and concentration
can be well approximated as $\chi (t) / c(t) \simeq 0.67(2)$ on the
time range indicated in Fig.~\ref{fig:marw}. This ratio agrees very well
with the theoretical prediction ($2/3$) obtained by Lee \cite{Lee}
using renormalization group technique.

\begin{figure}
\centerline{\epsfxsize=7.5cm
            \epsfbox{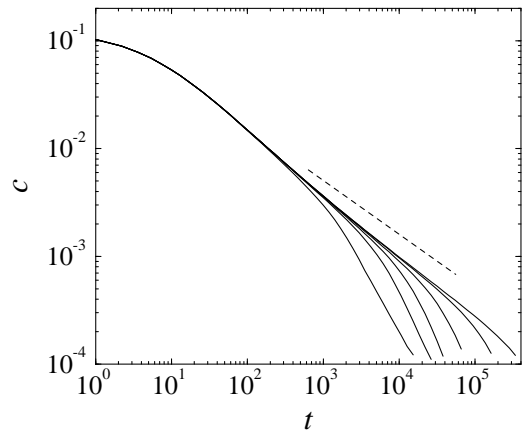}
            \vspace*{2mm}   }
\caption{Time dependence of concentration of particles (and antiparticles)
in the second model at zero branching rate for different system sizes
($L=100$, 200, 300, 500, 1000 and 2000 from left to right). The dashed
line (slope $-0.5$) indicates the theoretical power law behavior.}
\label{fig:paparw}
\end{figure}

For the second model the decrease of concentration follows a different
behavior at the critical point. Indeed when $p=0$ we get the diffusion
limited surface reaction A+B $\to \emptyset$ case, already studied
by several authors \cite{TW,CDC,LC}. The concentrations decrease as
$t^{-d/4}$ if the dimension $d$ is lower than 4 (the upper critical
dimension for this system). Compared to the former case, the results
depend more strikingly on size, as shown in a log-log plot
(Fig.~\ref{fig:paparw}). The sharp decrease in $c$ is a consequence of
the extinction whose probability is higher for smaller systems.
Our simulation confirms the
mentioned power law behavior in the limit $L \to \infty$ as indicated 
in Fig.~\ref{fig:paparw}. The extent of this behavior for the largest
system is illustrated in the figure  by the dashed line. 
The fluctuations are almost constant during this scaling regime
($\chi(t) \simeq 0.009(1)$ for $L=2000$ and $600<t<60000$). 

Systematic and extensive MC simulations have been performed to study
the average concentration and its fluctuation in the steady state
(reached after some thermalization) for finite branching rates.
For the smallest $p$ values both the thermalization and sampling
times were longer than $10^5$ ($10^6$) MCS for the first (second)
model. These simulations were repeated 20 times to suppress the
undesired effects of long time fluctuations.

\begin{figure}
\centerline{\epsfxsize=7.5cm
            \epsfbox{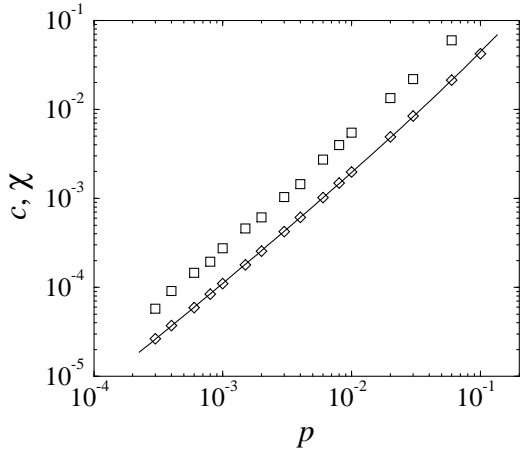}
            \vspace*{2mm}   }
\caption{Log-log plot of the particle concentration (diamonds) and its
fluctuation (squares) as a function of branching rate for the
first model. The solid line indicates the fitted curve described in
the text.}
\label{fig:mbarw}
\end{figure}

The results for the first model are summarized in Fig.~\ref{fig:mbarw}.
At first glance, the MC data for the concentration (diamonds in
the log-log plot) indicates a power law behavior, namely
$c(p) \propto p^{\beta}$ with $\beta=1.276$, similar to what was found
by Takayasu and Tretyakov\cite{TT}. The careful reader can, however,
observe a definite deviation from this behavior (positive curvature)
whose magnitude exceeds our statistical error.

Taking the logarithmic corrections into account Cardy and T\"auber have
suggested \cite{CT96} that the leading term of the $c(p)$ function is
proportional to $p/\ln ^2 (p)$. This function does not fit adequatly
 the present MC data; however, an excellent fitting is found if we use 
 $c(p)=p/[A + B \ln (p) + C \ln^2 (p)]$. This function
is represented by a solid line in Fig.~\ref{fig:mbarw} for
$A=0.4098$, $B=-0.5825$ and $C=0.09601$.
Notice that this formula confirms the theoretical prediction
in the limit $p \to 0$.

\begin{figure}
\centerline{\epsfxsize=7.5cm
            \epsfbox{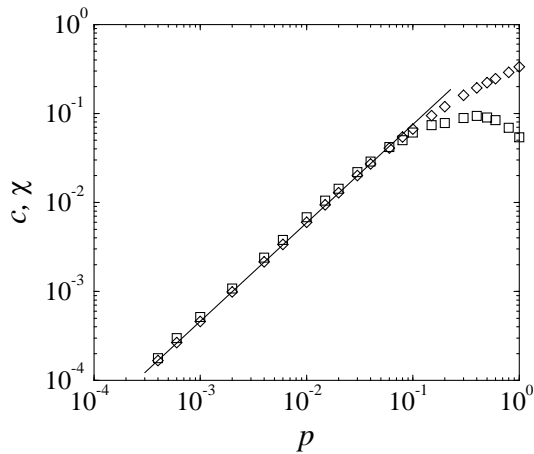}
            \vspace*{2mm}   }
\caption{Particle and antiparticle concentration (diamonds) and its
fluctuation (squares) {\it vs.} branching rate for the second model.
The fitted asymptotic power law behavior is indicated by the straight
line (slope $1.11$).}
\label{fig:papbarw}
\end{figure}

Figure \ref{fig:mbarw} indicates clearly that the concentration
fluctuation is proportional to the concentration itself in the $p$
region we have studied. These quantities
satisfy the relation $\chi (p) / c(p) = 2.5(2)$ within the statistical
error.

Similar investigations have been performed for the second model.
Figure~\ref{fig:papbarw} demonstrates clearly that the MC data tends
towards a power law for both the concentration ($c$) and its
fluctuation ($\chi$) at small values of branching $p$. 
Fitting the function $c(p)=A p^{\beta}$ to our numerical data we have
obtained that $\beta = 1.11(1)$. The ratio of the fluctuation to
the concentration is smaller than found for the first model,
namely, $\chi (p) / c (p) = 1.10(5)$.

In summary, we have studied and compared two simple models of
BARWs on a square lattice with parity conservation. In contrast
with the first model -- where there is only one type of particles --
the second model has particles and antiparticles annihilating
(only) each other when they meet. This distinction has caused
significant differences between their behavior at the critical
point (no branching) as well as in the stationary states for finite
branching rate. For the first model, our MC simulations have
justified the appearance of the logarithmic corrections predicted
theoretically by Lee and by Cardy and T\"auber. On the contrary, we have
observed power law behavior in the second model. Surprisingly, the
fluctuations decrease with $p$ -- more precisely, the $p$ dependence
of the fluctuation is found to be proportional to the concentration
for both models. The significant differences between the behavior of
the present models imply the possibility to find other two-dimensional
systems whose critical behavior does not belong to the directed
percolation universality class. The belief that in these systems
parity conservation is sufficient to determine the universality class is
probably also to be questioned. 

\acknowledgements
We thank U. T\"auber for his valuable comments.
We have benefited from discussions with J. F. Mendes and M. C. Marques.
G. S. acknowledges a senior research fellowship from PRAXIS (Portugal).
Supports from NATO (CRG-970332), PRAXIS (project PRAXIS/2/2.1/Fis/299/94)
and the Hungarian National Research Fund (T-16734) are also acknowledged.


\begin{references}
\bibitem{notation}In references \cite{TT} and \cite{J93}, the branching
probability is denoted by $1-p$.
\end{references}
\end{document}